\documentclass[a4paper,twocolumn,english,aps]{revtex4-1}

\usepackage[T1]{fontenc}
\usepackage[utf8]{inputenc}
\usepackage[english]{babel}
\usepackage{amsmath,amsfonts,amsthm,amssymb}
\usepackage{mathtools}
\usepackage{microtype}
\usepackage{txfonts}
\usepackage{graphicx}
\usepackage{xcolor}
\usepackage{bm}

\usepackage{siunitx}

\definecolor{citegreen}{rgb}{0.0, 0.5, 0.0}
\definecolor{linkCC}{RGB}{51,102,204}

\usepackage{hyperref}
\hypersetup{
    colorlinks,
    linkcolor={linkCC},
    citecolor={citegreen},
    urlcolor={blue!75!black}
}

\begin{document}

\title{Magnetoplasmons in monolayer black phosphorus structures}

\author{Yun~You$^{1,2,3}$, P.~A.~D.~Gon\c{c}alves$^{3,4,5}$, Linfang~Shen$^6$, Martijn~Wubs$^{3,4}$, Xiaohua~Deng$^{2}$, Sanshui~Xiao$^{3,4}$}
\email{saxi@fotonik.dtu.dk}

\affiliation{$^{1}$College of Material Science and Engineering, Nanchang University, Nanchang, 330031, China}
\affiliation{$^{2}$Institute of Space Science and Technology, Nanchang University, Nanchang, 330031, China}
\affiliation{$^{3}$Department of Photonics Engineering, Technical University of Denmark, DK-2800 Kgs. Lyngby, Denmark}
\affiliation{$^{4}$Center for Nanostructured Graphene, Technical University of Denmark, DK-2800 Kgs. Lyngby, Denmark}
\affiliation{$^{5}$Center for Nano Optics, University of Southern Denmark, Campusvej 55, DK-5230 Odense M, Denmark}
\affiliation{$^{6}$Department of Applied Physics, Zhejiang University of Technology, Hangzhou 310023, China}

\begin{abstract}
Two-dimensional materials supporting deep-subwavelength plasmonic modes can also exhibit strong magneto-optical responses.
Here, we theoretically investigate magnetoplasmons (MPs) in monolayer black phosphorus (BP) structures under moderate static magnetic fields.
We consider three different structures, namely, a continuous BP monolayer, an edge formed by a semi-infinite sheet, and finally, a triangular wedge configuration.
Each of these structures shows strongly anisotropic magneto-optical responses induced both by the external magnetic field and by the intrinsic anisotropy of the BP lattice.
Starting from the magneto-optical conductivity of a single-layer of BP, we derive the dispersion relation of the MPs in the considered geometries, using a combination of analytical, semi-analytical, and numerical methods.
We fully characterize the MP dispersions and the properties of the corresponding field distributions, and we show that these structures sustain strongly anisotropic subwavelength modes that are highly tunable.
Our results demonstrate that MPs in monolayer BP, with its inherent lattice anisotropy as well as magnetically induced anisotropy, hold potential for tunable anisotropic materials operating below the diffraction limit, thereby paving the way for tailored nanophotonic devices at the nanoscale.
\end{abstract}

\maketitle

\begin{figure}[t]
\centering
\includegraphics[width=\columnwidth]{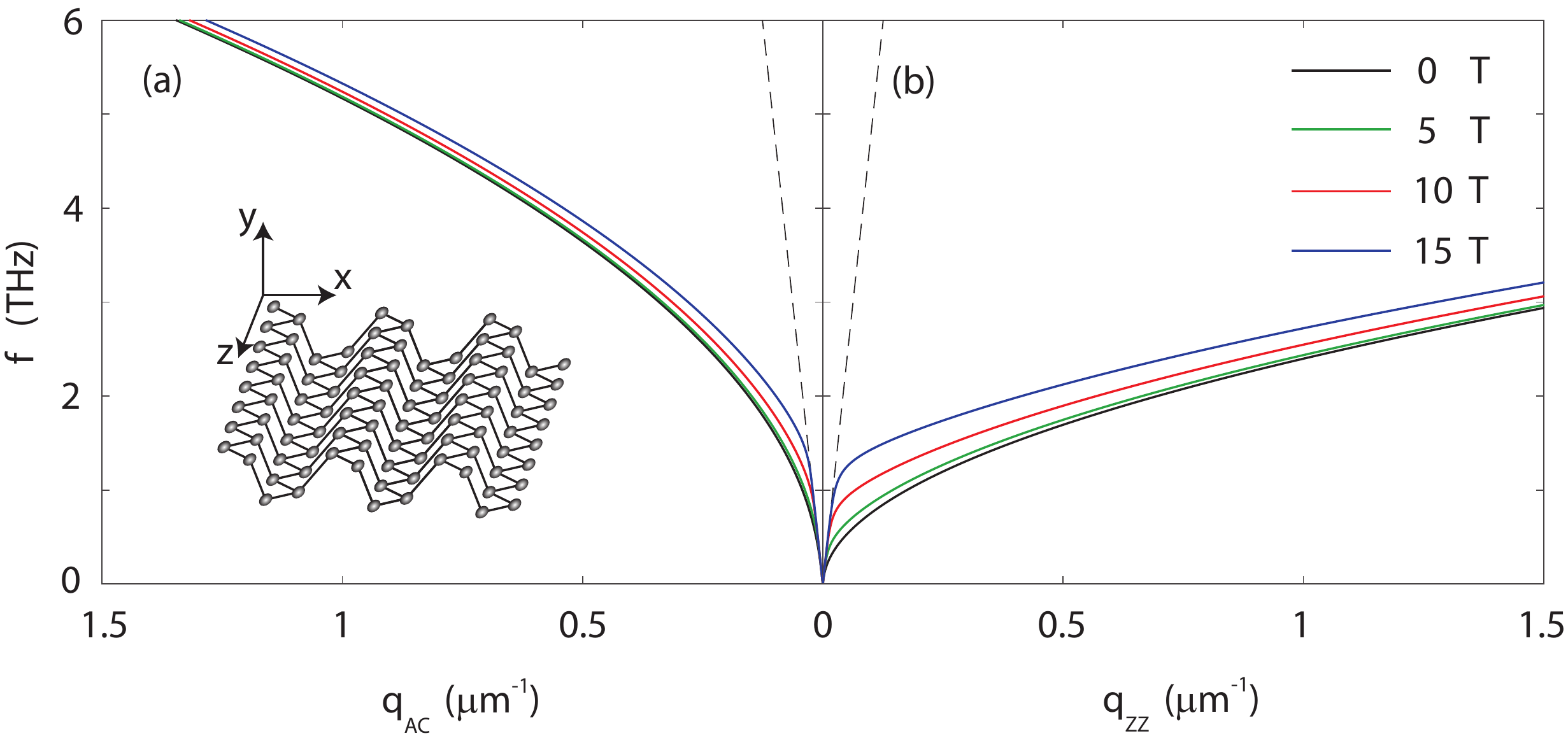}
\caption{Dispersion relation of magnetoplasmons in a continuous BP monolayer.
The perpendicular magnetic field is applied along the positive $y$-axis.
We consider propagation along both the (a) armchair and (b) zigzag crystallographic directions.
The dashed lines indicate the light-line.
The inset depicts a BP monolayer with the armchair and zigzag directions along the $x$- and $z$-axis, respectively.}
\label{fig:plane_plasmon}
\end{figure}

Surface plasmons~\cite{GoncalvesPeres_book}, collective oscillations of charge-carriers in conductors or doped semiconductors, possess a remarkable ability to confine optical fields below the diffraction limit~\cite{gramotnev2010plasmonics}, and thus hold a great potential to bridge the gap between electronics and photonics~\cite{ozbay2006plasmonics}.
When a static magnetic field is applied to a plasmonic material, the material's charge-carriers are affected by the Lorentz force.
The corresponding magneto-optical response then gives rise to surface magnetoplasmons (MPs).
Compared with surface plasmons, MPs provide extra tunability, since the plasmonic properties can be tailored by the applied magnetic field even for a fixed structure, geometry, and material.
In addition, MPs typically exhibit nonreciprocity, which may be advantageous for routing electromagnetic radiation in the subwavelength regime.
Indeed, with appropriate designs, unidirectional propagation of electromagnetic fields can be engineered~\cite{haldane2008possible,shen2015backscattering} by taking advantage of surface MPs.
External magnetic fields can also produce Landau levels and other quantization effects in graphene nanoribbons and phosphorene~\cite{Chung2016,Wu2017,PRB92_2015}.
Although surface plasmons exhibit many alluring properties, these excitations typically suffer from short lifetimes owing to the inherent losses in metals, which may hinder further applications.
The emergence of two-dimensional (2D) materials over the last decade---spearheaded by graphene---has attracted a vast amount of interest in photonics, as it has the potential to overcome the shortcomings of traditional metal-based plasmonics~\cite{GoncalvesPeres_book,koppens2011graphene,Chung2016,low2017polaritons}.
Graphene can sustain long-lived plasmons~\cite{koppens2011graphene,GoncalvesPeres_book,xiao2016graphene,low2017polaritons}, and 2D plasmonic materials can confine light in deep subwavelength regimes as a result of their intrinsic 2D property and plasmon dispersion.
Very recently, Iranzo \emph{et al.}~\cite{iranzo2018probing} reported extreme plasmon confinement based on a layered structure consisting of graphene, hexagonal boron nitride, and a metal grating to squeeze light within the length of a single layer.
Still, and despite its remarkable electronic and optical properties, realizing graphene-based devices for high-quality field-effect transistors is challenging due to the absence of a bandgap.
In this context, monolayer black phosphorus (BP) can provide a moderate bandgap, as well as high electronic mobility~\cite{xia2014two}.
Single-layer BP has a puckered crystal structure of phosphorus atoms that gives rise to an in-plane anisotropy with distinct electron (or hole) effective masses along the high-symmetry armchair (AC) and zigzag (ZZ) directions---see the inset in Fig.~\ref{fig:plane_plasmon}.
Its large anisotropy has been demonstrated both by Raman scattering spectra and photoluminescence measurements~\cite{zhang2014extraordinary}.
Recent research has also shown that BP-based field-effect transistors can exhibit outstanding on/off ratios and relatively high cut-off frequencies at the same time~\cite{li2014black}.
Other potential applications based on BP, such as a heterojunction $p$--$n$ diode~\cite{deng2014black}, photovoltaic devices~\cite{buscema2014photovoltaic}, or high responsivity photodetectors~\cite{engel2014black}, have also been demonstrated.

In this Letter, we theoretically investigate the properties of MPs supported by monolayer BP structures including a continuous BP sheet, an edge formed by a semi-infinite BP plane, and a triangularly shaped wedge made of single-layer BP.
In particular, we study the systems' magneto-optical response and determine the MPs dispersion relation, focusing on how the plasmonic properties and mode confinement are affected by the external magnetic field, depending on its strength and direction with respect to the BP plane(s).
Owing to BP's lattice anisotropy, we investigate the properties of its MPs both along the AC and ZZ directions.

Here, the monolayer BP is treated as a strictly 2D material without thickness, both in our analytical and numerical calculations.
The material's response is characterized by a frequency-dependent (surface) conductivity tensor that, in the absence of an external magnetic field, is of the following form~\cite{low2014plasmons}
\begin{equation}
\bm{\sigma} (\omega) = \frac{i n e^2}{\omega + i\gamma}
\begin{bmatrix}  m_{xx}^{-1} & 0 \\ 0 & m_{zz}^{-1} \end{bmatrix}
, \label{sigma}
\end{equation}
where $n$, $e$, $\omega$, and $\gamma/(2\pi)$ are, respectively, the carrier-density, elementary charge, angular frequency, and electronic scattering rate.
The diagonal terms, $m_{xx}^{-1}$ and $m_{zz}^{-1}$, in the magneto-optical conductivity tensor stand for different effective masses of the charge carriers along those directions, which can be oriented along the AC or ZZ crystal axes---see the inset in Fig.~\ref{fig:plane_plasmon}.
We note that, contrary to a three-dimensional (3D) bulk magneto-optical material, the monolayer BP only responds to the component of the magnetic field which is perpendicular to the material's surface.
In the presence of an applied static magnetic field, the conductivity tensor of the BP monolayer becomes
\begin{subequations} 
\begin{equation} 
\bm{\sigma} (\omega) =
 \begin{bmatrix}  \sigma_{xx} & \sigma_{xz} \\ -\sigma_{xz} & \sigma_{zz} \end{bmatrix} ,
\end{equation}
with
\begin{align}
 \sigma _{jj} &= \frac{i n e^2}{m_{jj}} 
 \frac{\omega + i\gamma}{ \left( \omega + i\gamma \right)^2 - \omega_c^2 } , \\
  \sigma _{jk} &= -\frac{n e^2}{m^{*}} 
 \frac{\omega_c}{ \left( \omega + i\gamma \right)^2 - \omega_c^2 } ,
\end{align}%
 \label{magnetic_sigma}%
\end{subequations}%
where $m^{*} = \sqrt{ m_{xx} m_{zz} }$ is the effective cyclotron mass, $\omega_c = e \mathbf{B}_{0} \cdot \mathbf{\hat{n}} / m^{*}$ is the cyclotron frequency, $\mathbf{B}_{0}$ is the (static) external magnetic field, and $\mathbf{\hat{n}}$ is the unit vector normal to the BP's surface.
Equation~(\ref{magnetic_sigma}) is derived within the framework of the Drude model~\cite{GoncalvesPeres_book}, considering the anisotropy of BP, as well as the influence of the external magnetic field.
With the anisotropic conductivity tensor derived above, we first study the MPs supported by a continuous BP monolayer sandwiched between two dielectrics.
In the case of a plasmonic 2D material experiencing a magnetic field, the presence of the latter effectively mixes the transverse electric (TE) and transverse magnetic (TM) modes.
The dispersion relation of the 2D MPs in the BP monolayer then follows from Maxwell's equations and corresponding boundary conditions~\cite{GoncalvesPeres_book}; assuming without loss of generality a MP propagating along the $x$-direction, its dispersion is determined from the solution of
\begin{equation}
\left( \frac{\varepsilon_1}{\kappa_1} + \frac{\varepsilon_2}{\kappa_2} + \frac{i \sigma _{\rm{xx}}}{\omega \varepsilon_0} \right) 
\left( \kappa_1 + \kappa_2 - i \omega \mu_0 \sigma {zz} \right) + \frac{\mu_0}{\varepsilon_0}{\sigma _{xz}}^2 = 0
. \label{dispersion}
\end{equation}
Here $\kappa_l = \sqrt{q^2 - \varepsilon_l \omega^2/c^2}$, with $q$ denoting the 2D MP wavevector, and $c$, $\epsilon_0$ and $\mu_0$ denoting the speed of light, permittivity and permeability in vacuum, respectively.
Moreover, $\varepsilon_l$ with $l=\{1,2\}$ label the relative permittivities of the cladding and substrate materials, respectively.
It should be noted that the dispersion relation for a MP propagating along the $z$-direction can be obtained by interchanging $\sigma_{xx}$ and $\sigma_{zz}$ in Eq.~(\ref{dispersion}).
Furthermore, notice that the term in the first parentheses is simply the TM mode dispersion of plasmons in an extended 2D material in the absence of a magnetic field~\cite{GoncalvesPeres_book}.
On the other hand, the term enclosed by the second parentheses corresponds to the dispersion of TE modes.
Finally, the last term represents the Hall conductivity induced by the external magnetic field in BP.
It is therefore clear that the presence of the static magnetic field mixes TM and TE modes, and, when the former is removed---which means that the Hall conductivity is zero---Eq.~(\ref{dispersion}) factorizes into the TM and TE plasmon dispersions [and the conductivity in Eq.~(\ref{magnetic_sigma}) reduces to the original tensor figuring in Eq.~(\ref{sigma})].

Figure~\ref{fig:plane_plasmon} shows the dispersion relations of MPs in an extended BP monolayer under different magnetic field strengths and assuming negligible losses; MP propagation along both the AC and ZZ directions is considered.
Throughout this Letter, we consider air to be the dielectric environment ($\varepsilon_1=\varepsilon_2=1$) and the BP's carrier-density to be 
$n = 10^{13}\, \mathrm{cm}^{-2}$.
For the electron effective masses, we take $m_{AC} = 0.15 m_0$ and $m_{ZZ} = 0.7 m_0$~\cite{low2014plasmons}, where $m_0$ is the electron rest mass.
\begin{figure*}[t]
\centering
\includegraphics[width=0.9\linewidth]{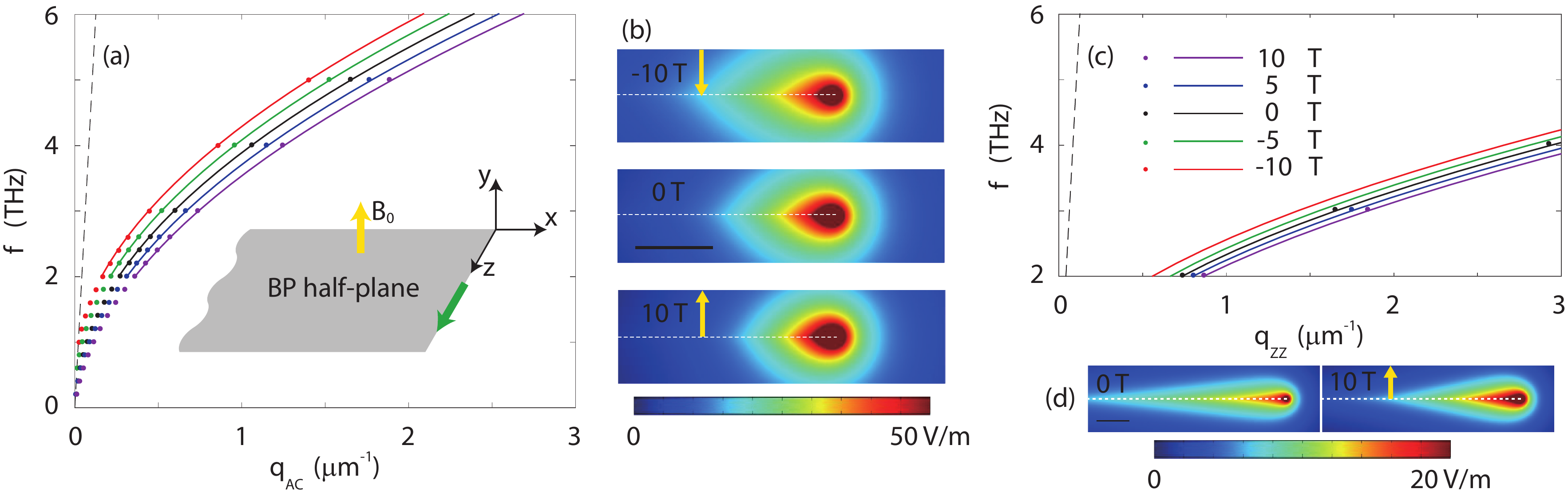}
\caption{Edge magnetoplasmons in a monolayer BP half-plane.
Dispersion of edge MPs propagating along the armchair (a) and zigzag (c) directions.
The modes propagate along the positive $z$-direction, and the static magnetic fields are applied along $\pm \mathbf{\hat{y}}$, with the BP monolayer lying in the $xz$-plane for $x<0$; see the inset in (a).
In both dispersion panels, the solid lines and the dots represent, respectively, the semi-analytical and numerical results.
The former are shown only within the validity of the nonretarded regime.
The light-line (black dashed lines) is also depicted.
(b) and (d) show electric-field amplitude distributions at \SI{4}{\tera\hertz} for armchair and zigzag directions, respectively.
The scale bar corresponds to \SI{0.5}{\micro\metre}.
The parameters are defined in the main text.}
\label{fig:edge_plasmon}
\end{figure*}
It is clear from Fig.~\ref{fig:plane_plasmon} that the 2D MPs dispersions lie close to the light-line for $\omega \lesssim \omega_c$ (as an example, $\omega_c = \SI{0.86}{\tera\hertz}$ for a $\SI{10}{\tesla}$ magnetic field), after which they start to depart from it in a similar fashion as the zero-field case.
Moreover, notice the strong anisotropic response: for a MP propagating along the BP's ZZ direction, where the electron effective mass is higher, the group velocity at a given frequency is correspondingly smaller in comparison to the AC case.
On the other hand, the effect of the magnetic field is stronger in the case where the MPs propagate along the heavier (ZZ) direction.

Edge and wedge structures made out of 2D materials can also sustain plasmonic excitations~\cite{GoncalvesPeres_book,Goncalves:2D_Nanoslits,gonccalves2016graphene,gonccalves2017universal}.
However, the nanostructuring of these beyond the simple, infinite and continuous 2D layer complicates the theoretical treatment, and \emph{fully analytical} solutions are not straightforward and easy to obtain.
Nevertheless, reliable and rigorous semi-analytical methods within the nonretarded limit (which often is the interesting regime for deep subwavelength plasmonics) can be employed.
In what follows, we use both semi-analytical and numerical methods to study quasi-one-dimensional MPs traveling along the edge of a semi-infinite BP monolayer.
The quasi-analytical model is based on (scalar) Green's functions, together with an orthogonal polynomial expansion; for details, we refer to Refs.~\cite{GoncalvesPeres_book,Goncalves:2D_Nanoslits}.
This method provides an accurate and reliable description of surface plasmons, supported by many different nanostructures and geometries based on 2D materials~\cite{GoncalvesPeres_book,Goncalves:2D_Nanoslits,gonccalves2016graphene}.
Our semi-analytical calculations are then augmented by full-wave numerical computations using the commercial software COMSOL Multiphysics based on the finite element method.
In the latter, the 2D BP monolayer is represented as a surface current, characterized by a conductivity tensor [recall Eq.~(\ref{magnetic_sigma})].
We stress that treating the monolayer as a surface current has advantages in terms of computational time and numerical accuracy when compared with the traditional volumetric method~\cite{Boltasseva:15}.
The dispersions of the edge MPs for both AC and ZZ edges are shown in Fig.~\ref{fig:edge_plasmon}, for different applied magnetic field strengths and orientations.
In both cases, we demonstrate that one may tune the MP dispersion by controlling the magnetic field strength.
Clearly, our semi-analytical model performs extremely well in the nonretarded regime, showing remarkable agreement with the numerical results.
In addition, notice that for perpendicular magnetic fields along the positive (negative) $y$-direction, a MP at a given frequency will exhibit higher (smaller) confinement than the corresponding zero-field edge plasmon.
This behavior can be understood by considering the Lorentz force: for a static magnetic field oriented along the positive $y$-direction, the charge-carriers are pushed towards the edge by the Lorentz force, which in turn results in a higher mode confinement~\cite{volkov1988edge} (and vice-versa for a magnetic field oriented along the negative $y$-direction).
The change of the mode confinement discussed here is consistent with the dispersion shift in Fig.~\ref{fig:edge_plasmon}.
We further stress that MPs propagating along a BP half-plane become nonreciprocal, as reciprocity is broken by the presence of the magnetic field.
In fact, it is interesting to note that flipping of the magnetic field, that is, from $y \rightarrow -y$, is equivalent to swapping of the propagating direction from along the positive $z$-direction to along the negative $z$-direction, and therefore (as Fig.~\ref{fig:edge_plasmon} shows) the dispersion for edge MPs propagating along the positive or negative $z$-direction is not identical.
Figures~\ref{fig:edge_plasmon}(b) and \ref{fig:edge_plasmon}(d) portray the electric field distributions of edge MPs with opposite magnetic field directions for the AC and ZZ propagating directions, respectively.
Note that MPs propagating along the ZZ direction have longer tails extending into the BP half-plane due to a stronger magnetic effect.
Although we have less numerical data for the ZZ case due to the time-consuming calculation process, the semi-analytical methods can provide reliable results of the dispersion of edge MPs in the nonretarded regime.
Finally, it should be mentioned that MPs traveling along the edge of a semi-infinite 2D sheet deliver higher field confinements when compared to its continuous sheet counterpart (cf. Figs.~\ref{fig:plane_plasmon} and ~\ref{fig:edge_plasmon}).

\begin{figure}[t]
\centering
\includegraphics[width=0.9\linewidth]{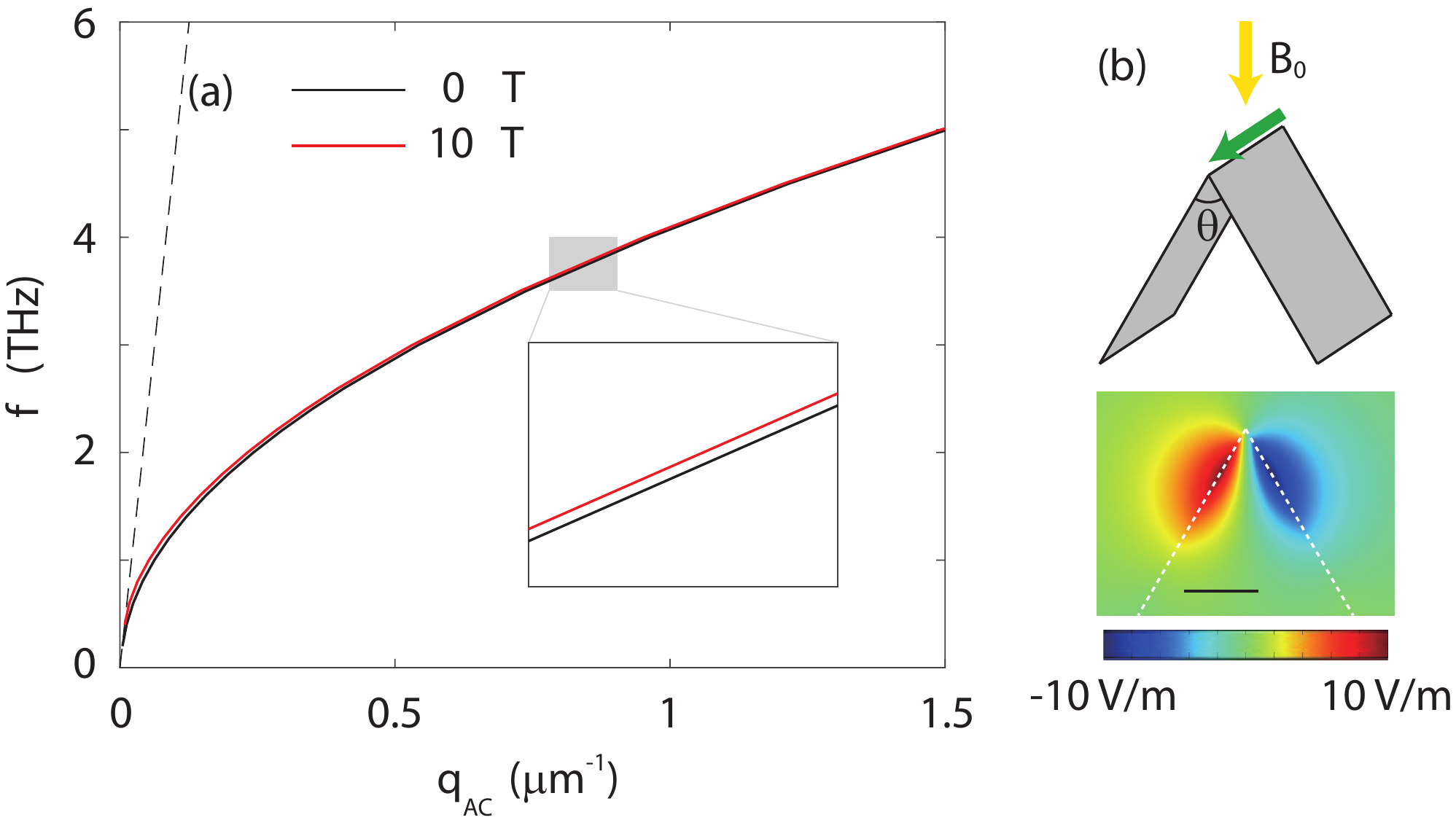}
\caption{(a) MP dispersion relation for a wedge of monolayer BP with and without an applied magnetic field (oriented along $-\mathbf{\hat{y}}$).
The angle formed by the wedge's apex is $60^{\circ}$, and the coordinate axes are defined as presented in the inset of Fig.~\ref{fig:edge_plasmon}(a).
(b) Sketch of the wedge structure consisting of BP monolayer.
Green and yellow arrows represent the directions of the propagating MPs and the external magnetic field, respectively.
Field distribution for $E_z$ component of the mode at $\SI{4}{\tera\hertz}$ under a $\SI{10}{\tesla}$ magnetic field.
The scale bar corresponds to \SI{1}{\micro\metre}.
The parameters are defined in the main text.}
\label{fig:wedge plasmon}
\end{figure}

After investigating MPs in flat structures of monolayer BP, we now consider a nonplanar geometry in which the BP monolayer is folded into a wedge structure forming a triangular channel~\cite{gonccalves2016graphene,gonccalves2017universal,wedge_experimental}.
Compared to the plane and edge structures, a wedge structure composed of a 2D material provides a flexible way to tune the plasmon mode by the external magnetic field because the two wedge arms exhibit different responses for a given external magnetic field.
It has been shown that plasmons in graphene wedges attain odd-symmetry modes (i.e., where the $E_z$ component has different signs with respect to the line bisecting the triangular cross section), which can be highly confined near the apex of the wedge~\cite{gonccalves2016graphene,gonccalves2017universal}.
Here, we now study this mode in the case of an anisotropic 2D material, a BP monolayer, subjected to a static magnetic field.
The corresponding wedge MP's dispersion is plotted in Fig.~\ref{fig:wedge plasmon}(a) for a wedge forming a $60^{\circ}$ angle.
It is apparent from the figure that in the presence of the magnetic field, the dispersion curve of the wedge MP moves upward, albeit the observed shift is smaller in comparison to the edge MP case under the same magnetic field strength.
This can be explained by the fact that only the perpendicular projection of the magnetic field with respect to the 2D surface, $B_0 \sin(\theta/2)$, drives the magneto-optical response of the 2D medium.
Figure~\ref{fig:wedge plasmon}(b) shows a sketch of the wedge structure and the field distribution of the $E_z$ component for MPs traveling along the AC direction with a $\SI{10}{\tesla}$ static magnetic field applied along the $-\mathbf{\hat{y}}$ direction.
The distribution of the $E_z$ component is not symmetric with respect to the bisecting line of the wedge due to the presence of the external magnetic field.
It is difficult to see the asymmetry in Fig.~\ref{fig:wedge plasmon} because the dispersion curve of the wedge MPs is close to its zero-field counterpart.
In the specific case discussed here, flipping the magnetic field direction will not make any difference to the dispersion relation.
This can be understood from the fact that only the distributions of charge-carriers of the two wedge arms are exchanged when the external magnetic field is swapped, as the external magnetic field is applied along the symmetry axis ($y$ axis) of the structure.

In conclusion, we have investigated the magneto-optical response and ensuing MPs supported in extended monolayer BP, edge MPs in a BP half-plane, and in a wedge BP structure.
We have determined the dispersion relations for MPs in a continuous BP sheet analytically and have employed a semi-analytical method to determine the dispersion relation of quasi-one-dimensional edge MPs in the half-plane case, which we then benchmarked against electrodynamic numerical calculations having obtained an outstanding agreement in the nonretarded limit.
Lastly, we have computed the wedge MP's dispersion using numerical means alone.
Our results demonstrate that the introduction of a static external magnetic field can be used to tune the MP's dispersion---both by varying the magnetic field strength and orientation---and the corresponding field confinement.
Furthermore, due to inherent anisotropy of BP, MPs propagating along the AC and ZZ differ significantly, with the former MPs attaining larger group velocities when compared to the latter.
In addition, we have shown that MPs in the BP edge exhibit nonreciprocal behavior owing to the presence of the magnetic field.
Finally, wedge MPs in which the magnetic field is applied along the line bisecting the triangular cross section show mode distortion in agreement with the Lorentz force induced by the external magnetic field.
Our results demonstrate that MPs in anisotropic 2D materials provide enhanced tunability together with nonreciprocity, which may be explored for controlling electromagnetic radiation below the diffraction limit, thereby providing new opportunities for designing novel tunable plasmonic devices.

\hfill\\
\noindent
\textbf{Funding.} This work was partly supported by the National Natural Science Foundation of China (NSFC) (61372005, 41331070), and Innovation Fund Designated for Graduate Students of Nanchang University (CX2017001). The Center for Nanostructured Graphene is sponsored by the Danish National Research Foundation (Project No. DNRF103).

\bibliography{reference}

\end{document}